\documentclass[superscriptaddress,
amsmath,amssymb,
aps,
prl,
floatfix,
]{revtex4-1}
\usepackage{mwe}
\usepackage{bm}
\usepackage{graphicx}
\usepackage[normalem]{ulem}
\usepackage{graphicx}
\usepackage{epstopdf}
\usepackage{hyperref}
\usepackage[normalem]{ulem}
\usepackage{xcolor}
\usepackage{bm}
\usepackage{url}
\usepackage{lipsum}
\usepackage{babel}
\usepackage{float}
\usepackage{lineno}

\newcommand{\tc}{T_c}
\newcommand{\Tc}{$\tc$}

\usepackage{bm}

\newcommand{\TS}{$\mathrm{TaS_{2}}$}
\newcommand{\TST}{1T-$\mathrm{TaS_{2}}$}
\newcommand{\TSH}{2H-$\mathrm{TaS_{2}}$}
\newcommand{\TSB}{4Hb-$\mathrm{TaS_{2}}$}	

\begin{document}

\title{Charge transfer and spin-valley locking in 4Hb-TaS$_2$}

\author{Avior Almoalem}
\affiliation{Physics Department, Technion-Israel Institute of Technology, Haifa 32000, Israel.}
\author{Roni Gofman}
\affiliation{Physics Department, Technion-Israel Institute of Technology, Haifa 32000, Israel.}
\author{Yuval Nitzav}
\affiliation{Physics Department, Technion-Israel Institute of Technology, Haifa 32000, Israel.}
\author{Ilay Mangel}
\affiliation{Physics Department, Technion-Israel Institute of Technology, Haifa 32000, Israel.}
\author{Irena Feldman}
\affiliation{Physics Department, Technion-Israel Institute of Technology, Haifa 32000, Israel.}
\author{Jahyun Koo}
\affiliation{Department of Condensed Matter Physics, Weizmann Institute of Science, Rehovot 7610001, Israel}
\author{Federico Mazzola}
\affiliation{Istituto Officina dei Materiali (IOM)-CNR, Area Science Park, S.S.14, Km 163.5, 34149 Trieste, Italy}
\author{Jun Fujii}
\affiliation{Istituto Officina dei Materiali (IOM)-CNR, Area Science Park, S.S.14, Km 163.5, 34149 Trieste, Italy}
\author{Ivana Vobornik}
\affiliation{Istituto Officina dei Materiali (IOM)-CNR, Area Science Park, S.S.14, Km 163.5, 34149 Trieste, Italy}
\author{J. Sánchez-Barriga}
\affiliation{Helmholtz-Zentrum Berlin für Materialien und Energie, BESSY II, Albert-Einstein-Strasse 15, 12489 Berlin, Germany}
\affiliation{IMDEA Nanoscience, C. Faraday, 9, Fuencarral-El Pardo, 28049 Madrid, Spain}
\author{Oliver J. Clark }
\affiliation{Helmholtz-Zentrum Berlin für Materialien und Energie, BESSY II, Albert-Einstein-Strasse 15, 12489 Berlin, Germany}
\author{Nicholas Clark Plumb}
\affiliation{Photon Science Division, Paul Scherrer Institut, CH-5232 Villigen PSI, Switzerland}
\author{Ming Shi}
\affiliation{Photon Science Division, Paul Scherrer Institut, CH-5232 Villigen PSI, Switzerland}
\author{Binghai Yan}
\affiliation{Department of Condensed Matter Physics, Weizmann Institute of Science, Rehovot 7610001, Israel}
\author{Amit Kanigel}
\affiliation{Physics Department, Technion-Israel Institute of Technology, Haifa 32000, Israel.}

\begin{abstract}
\TSB  ~is a superconductor that exhibits unique characteristics such as time-reversal symmetry breaking, hidden magnetic memory, and topological edge modes. It is a naturally occurring heterostructure comprising of alternating layers of 1H-TaS$_2$ and 1T-TaS$_2$. The former is a well-known superconductor, while the latter is a correlated insulator with a possible non-trivial magnetic ground state.
In this study, we use angle resolved photoemission spectroscopy to investigate the normal state electronic structure  of this unconventional superconductor. 
Our findings reveal that the band structure of \TSB\ fundamentally differs from that of its constituent materials. Specifically, we observe a significant charge transfer from the 1T layers to the 1H layers that drives the 1T layers away from half-filling.  In addition, we find a substantial reduction in inter-layer coupling in \TSB\ compared to the coupling in \TSH\ that results in a pronounced spin-valley locking within \TSB.

\end{abstract}

\maketitle

\section{Introduction}

Over the past few years, the field of quantum condensed matter research has undergone a significant revolution, thanks to breakthroughs in stacking atomically thin materials with diverse properties and at arbitrary relative angles~\cite{novoselov20162d}. 

These heterostructures exhibit a variety of  interacting phases, including superconductivity (SC)~\cite{cao2018unconventional} , correlated insulators~\cite{cao2018correlated}, structural and electronic ferroelectrics ~\cite{vizner2021interfacial}, electronic nematicity ~\cite{cao2021nematicity}, and magnetism~\cite{chen2020tunable}.

Alternate stacking structures of TaS$_2$ have also drawn significant attention. The natural occurring heterostructure, \TSB, formed by alternate stacking of \TST\ and 1H-\TS~\cite{DiSalvo1973preparation,ribak2017gapless,Nayak2021evidence,persky2022magnetic} is a superconductor with unusual properties that led some of us to propose that it is a chiral superconductor \cite{Ribak2020,Nayak2021evidence,persky2022magnetic,almoalem2022evidence}.
In contrast, bi-layers made by growing mono-layers of \TST\ and 1H-\TS, are not superconducting and instead show Kondo physics~\cite{vavno2021artificial} or doped Mott physics \cite{crippa2023heterogeneous}.

The charge density wave (CDW) reconstruction in \TST\ results in a state consisting of a super-lattice, where every 13 atoms form a star-of-David cell with one localized electron. In bulk \TST\ this leads to a Mott transition and possibly to a quantum spin liquid \cite{law20171t, ribak2017gapless}.

These localized states are believed to play a crucial role both in \TSB\ and in the bi-layers. In \TSB, it was suggested that the proximity of the strongly correlated 1T layers to the superconducting 1H layers is responsible for the observed time-reversal symmetry breaking \cite{Ribak2020,liu2023magnetization,konig2023type}. In the artificial bi-layers, the localized moments on the 1T layers are Kondo screened by the itinerant electrons in the 1H layers \cite{vavno2021artificial}.

A remaining open question is how the proximity between the 1T and the 1H layers alters their properties, specifically, the extent to which the Mott physics, believed to govern the physics of \TST\, plays a role in \TSB.
In this paper, we address this question using angle resolved photoemission spectroscopy (ARPES), comparing in detail the electronic structure of \TSH\ and \TSB.

\section{Results}

\subsection{Charge Transfer In 4Hb-TaS$_{2}$}

\begin{figure*}[!ht]
	\centering
	\includegraphics[width=1\linewidth]{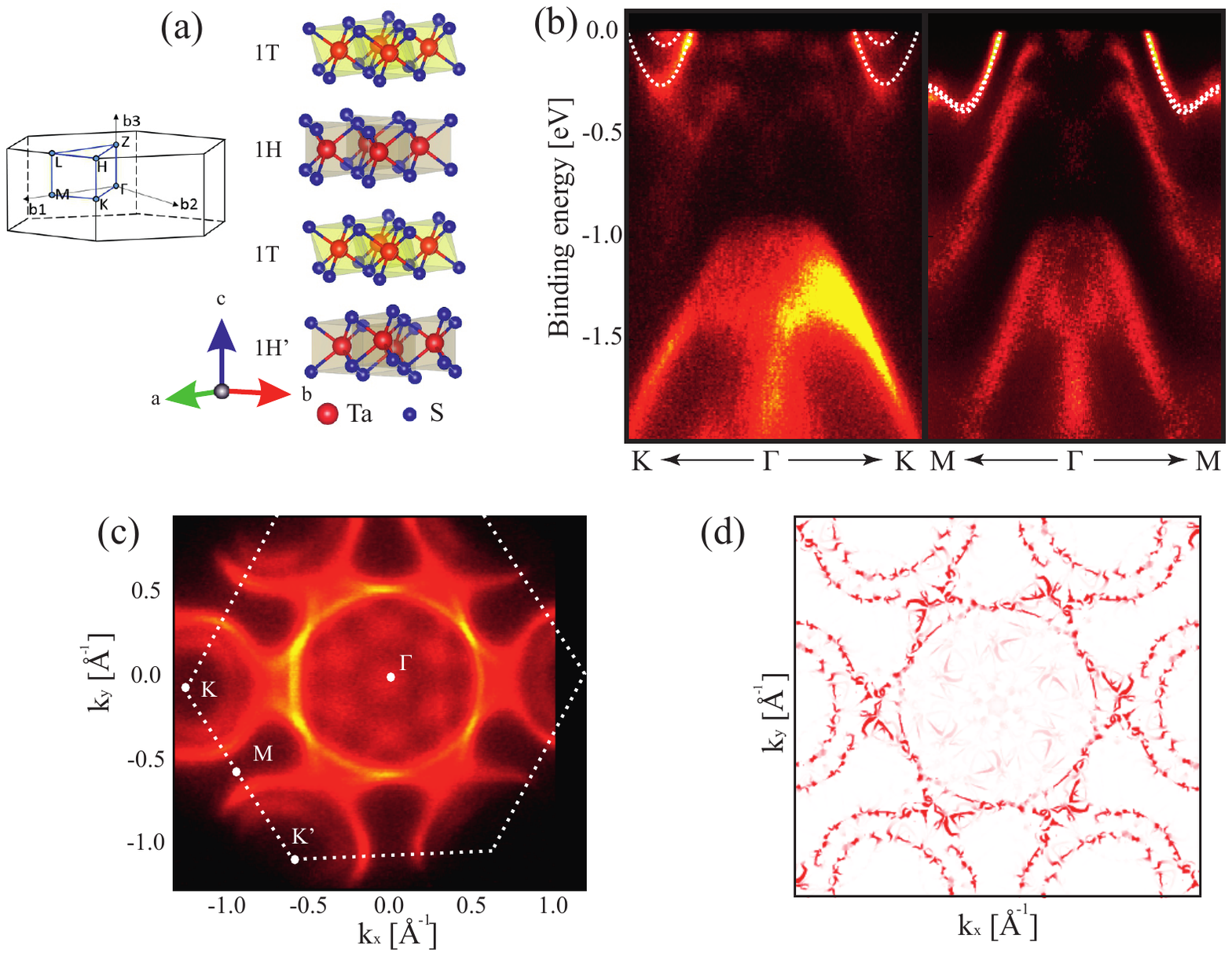}
	\caption{\textbf{Band Structure and Fermi surface of \TSB.} \textbf{(a)} The unit cell of \TSB\ with the 1T and 1H layers. The 1H and 1H' layers are rotated by 180$^\circ$ degrees. Brillouin zone of the 3D crystal structure showing the high symmetry points. b$_1$, b$_2$ and b$_3$ are reciprocal lattice vectors. \textbf{(b)} ARPES spectra taken with $\hbar\nu$=102~eV. The spectra is taken along high symmetry lines of the Brillouin zone.  Dashed lines are DFT calculations of the 1H-layers derived bands in 4Hb-TaS$_2$, done on a slab of 10 unit cells. Other bands are different from bands in the  bulk 2H and 1T polymorphs   \cite{Lahoud2014emergence}.  Shallow electron pockets are visible at the $\Gamma$ point matching qualitatively the  DFT results.  \textbf{(c)}  Fermi-surface of \TSB. The data is in good agreement with the DFT results. The 2D hexagonal Brillouin zone is shown with white dashed lines. \textbf{(d)} Fermi surface from a DFT calculation on a 1T/1H bi-layer. The 1T CDW ($\sqrt{13}\times\sqrt{13}$) was included in calculations and the Fermi surface was unfolded to the original Brillouin zone for comparison with experiments. The small Fermi pockets produced by the shallow bands are clearly seen. }
	\label{Fig1}
\end{figure*}

Both 4Hb and 2H polymorphs of \TS\ belong to the $\mathrm{P6_3/mmc}$ hexagonal space group, the unit cell of \TSB\ is shown in Fig.~\ref{Fig1}(a). \TSB\ undergoes a CDW transition at $T=315$~K and a sharp SC transition at \Tc=2.7~K\cite{DiSalvo1973preparation,Kim1995atomic,Ribak2020}, while \TSH\ has a CDW transition at 75K and \Tc\ of 0.8K~\cite{DiSalvo1973preparation}. Most of the intensity  at the Fermi level in both 2H and 4Hb is from the Ta atoms in 1H layers which have a non-centrosymmetric structure ~\cite{DiSalvo1973preparation}. Both polymorphs contain two 1H layers stacked with a relative 180$^{\circ}$ rotation in-plane, resulting in an overall inversion-symmetric structure~\cite{Ribak2020,goryo2012possible}.


ARPES spectra along the high symmetry lines of \TSB, $\Gamma$-K and $\Gamma$-M, are shown in Fig.~\ref{Fig1}(b). Similar data from \TSH\ is shown in Fig.~S1 of the supplementary material. Comparing the ARPES data for both polymorphs reveals that the bands derived from the 1H layers are similar in both structures but not identical.

\begin{figure}[!ht]
	\centering
	\includegraphics[trim={0cm 0cm 0cm 0cm},clip, width=0.5\linewidth]{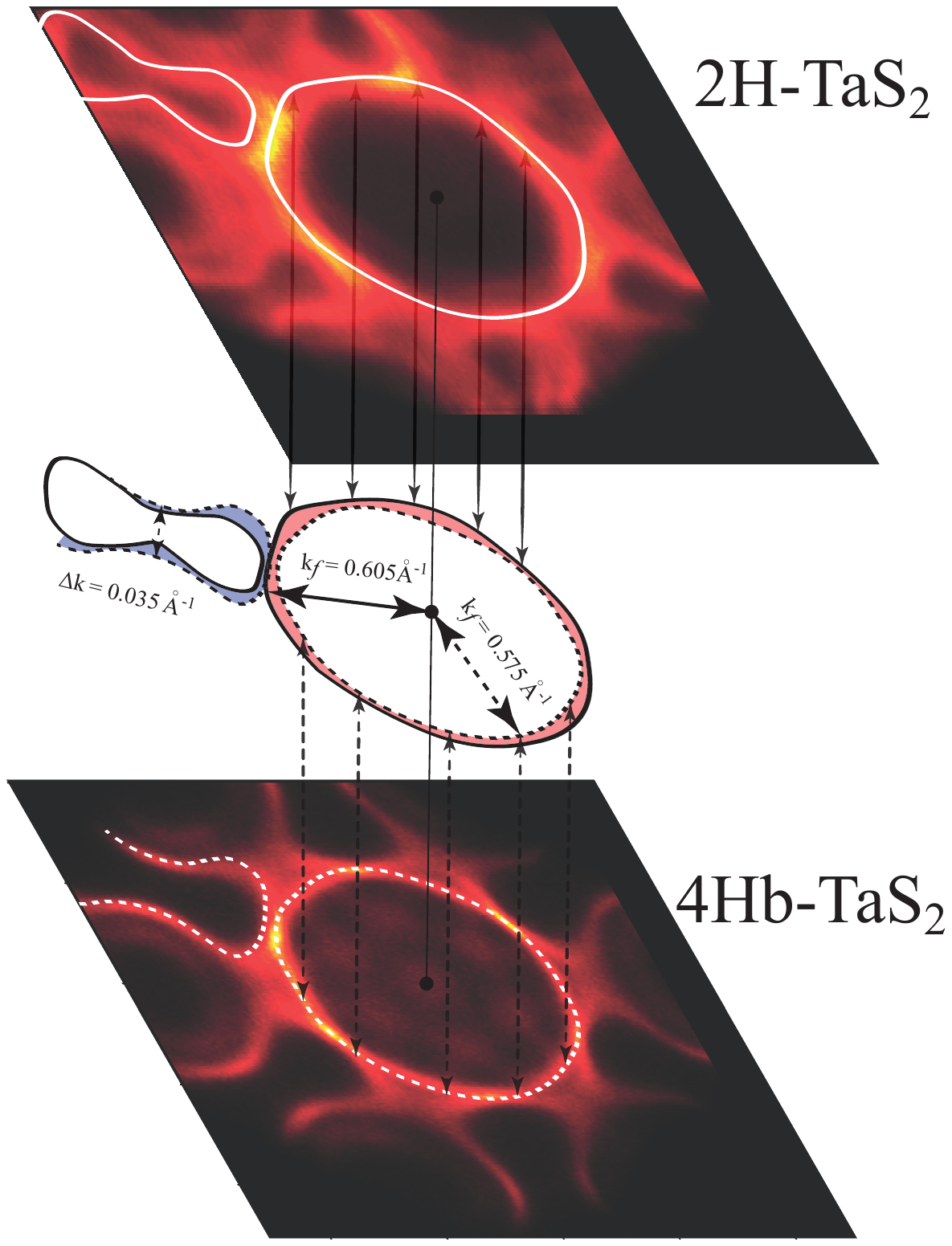}
	\caption{\textbf{Charge transfer in \TSB.} Fermi surface contours extracted from ARPES spectra of \TSH\ (upper panel) and \TSB\ (lower panel) overlapped. The dashed lines correspond to \TSB, and the solid lines correspond to \TSH, the lines were drawn by following the maximum intensity at the Fermi level. In the middle panel, red filling shows the contracting hole pocket around the $\Gamma$ point, and the blue filling shows the expanding "dog-bone" electron pockets, around the M point. k$_F$ measured from the M-point in the narrow part "dog-bone" is 0.323~\AA$^{-1}$ and 0.358~\AA$^{-1}$ for \TSH, and \TSB\ respectively. The change in k$_F$ of the K-point pocket is too small to show in the figure.  The estimated errors in k$_F$ of the hole pocket and the "dog-bone" are 0.008~\AA$^{-1}$ and 0.005~\AA$^{-1}$ respectively. Overall, the differences agrees with the DFT results of the 1H/1T bi-layer.  Data for both samples was measured close to k$_Z$=0 (92eV for \TSH\ and 102eV for \TSB).}  
	\label{Fig3}
\end{figure}

The inner pocket along the K-M-K' line, derived from the 1H layer, is $\sim100$meV deeper compared to \TSH. Additionally, there are other bands originating from the 1T layers, these bands are notably different from the bands in bulk \TST\ \cite{Lahoud2014emergence}. 
In particular, there are several shallow pockets with Fermi energies of about 50meV around the $\Gamma$ point.

The measured Fermi-surface is shown in Fig. \ref{Fig1}c, and the  Fermi-surface calculated using density-functional theory (DFT) is shown in Fig.~\ref{Fig1}d. The agreement between the DFT and the ARPES data is remarkable. The new 
bands form a rather complicated set of small pockets around $\Gamma$ that seem to have a chiral structure, reminiscent of the chirality found in the charge density wave (CDW) in bulk 1T-TaS$_2$ \cite{Chiral_1T}. The ARPES data is in agreement with the STM data from the same sample\cite{nayak2023first}, which identifies a band located just above the Fermi level with only a weak density of states crossing the Fermi level.  The role of these new bands in superconductivity is not clear. It has been suggested that Kondo physics seen in 4Hb \cite{nayak2023first,shen2022coexistence} and in 1T/1H bilayers \cite{vavno2021artificial} could be related to these narrow bands. 

Other bands along the $\Gamma$-K line, suggested to originate from the 1T layer by DFT, are similar  to bands in bulk \TST. 

DFT finds a significant charge transfer from the 1T to the 1H layers. The charge transfer depends strongly on the inter-layer distance\cite{crippa2023heterogeneous}. For a spacing of 5.9 \AA\ it is found that the entire localized electron on the star-of-David cluster is transferred to the 1H layers \cite{Nayak2021evidence}.

The ARPES data show clear evidence of charge transfer. First, the pocket along  the K-M-K' line is deeper in \TSB\ by about 100~meV compared to \TSH\ (see supplementary Fig.~S2 and Fig.~S3) . The agreement between the DFT and the ARPES data further supports the observation of the charge transfer. To directly measure the charge transfer, we compare the Fermi-surface area of \TSB\ and \TSH\ in Fig. \ref{Fig3}. We find that the hole pocket around $\Gamma$ of \TSB\ is smaller than in \TSH, while the "dog-bone"-shaped electron pocket is larger. 

\begin{figure*}[!ht]
	\centering
	\includegraphics[width=1\linewidth]{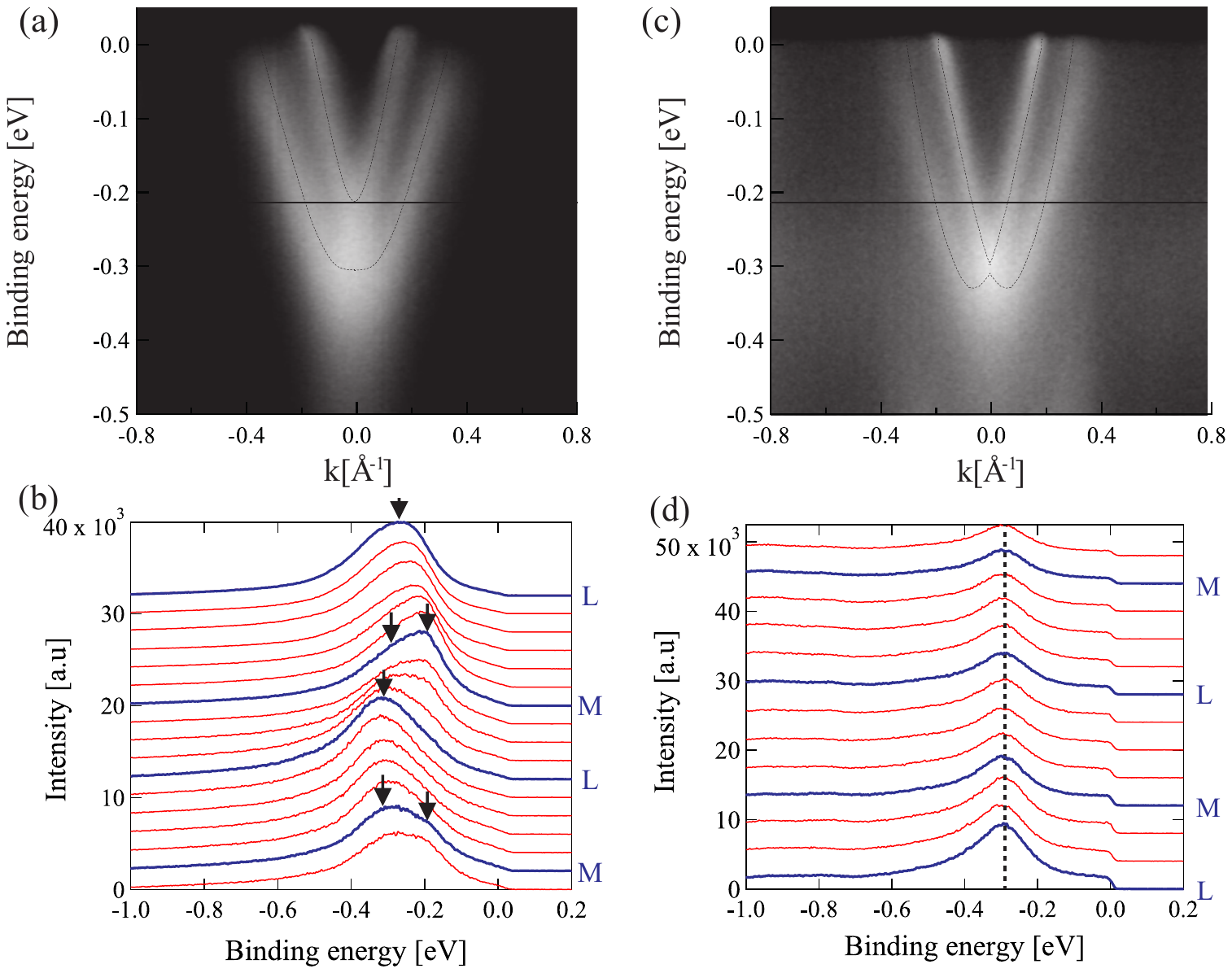}
	\caption{\textbf{Interlayer coupling.} \textbf{(a, c)} ARPES data  along the K-M-K' line for \TSH\ and \TSB, respectively. The solid lines are calculated surface bands from a 4Hb slab model. In \TSH\ there is a clear gap at the M-point which is a result of a significant inter-layer coupling.  In \TSB\ we find no gap in the ARPES data within our resolution (about 5meV), suggesting that the 1H layers are very weakly coupled.  The 2D nature of \TSB\ is also evident in the k$_z$ dispersion. In \textbf{(b,d)} we show the dispersion along the M-L line for \TSH\ and \TSB. The dispersion normal to the plane is measured by scanning the photon energy. In \TSH we find a band width of about 150 meV. The inter-layer coupling is maximal at the M-point and vanishes  at the L-point.   In \TSB, on the other hand, we find a completely flat band normal to planes, as expected from a purely 2D system. Arrows mark the energy of the bands at each k$_z$, and were obtained by fitting the EDCs to two Lorentzian functions. For \TSH\  we measured with photon energies between 46ev  and 82eV, and for \TSB\ with photon energies between 49eV and 66eV.}  
	\label{Fig2}
\end{figure*}

Quantitatively, by integrating the differences between the Fermi-surface areas, we estimate the total charge transfer to be $0.071\pm0.008$ electron per Ta atom, or $0.92\pm0.11$ per star-of-David (see supplementary material for details). It is worth noting that this estimate ignores the contribution of the new small pockets around $\Gamma$, but these seem to enclose a very small area. Nevertheless, the fact that these shallow bands do produce Fermi-pockets, suggests that although the charge transfer is significant the depletion of the 1T flat band is not complete. For the details of the charge transfer estimate, see Supplementary material.

\begin{figure*}[!ht]
	\centering
	\includegraphics[width=1\linewidth]{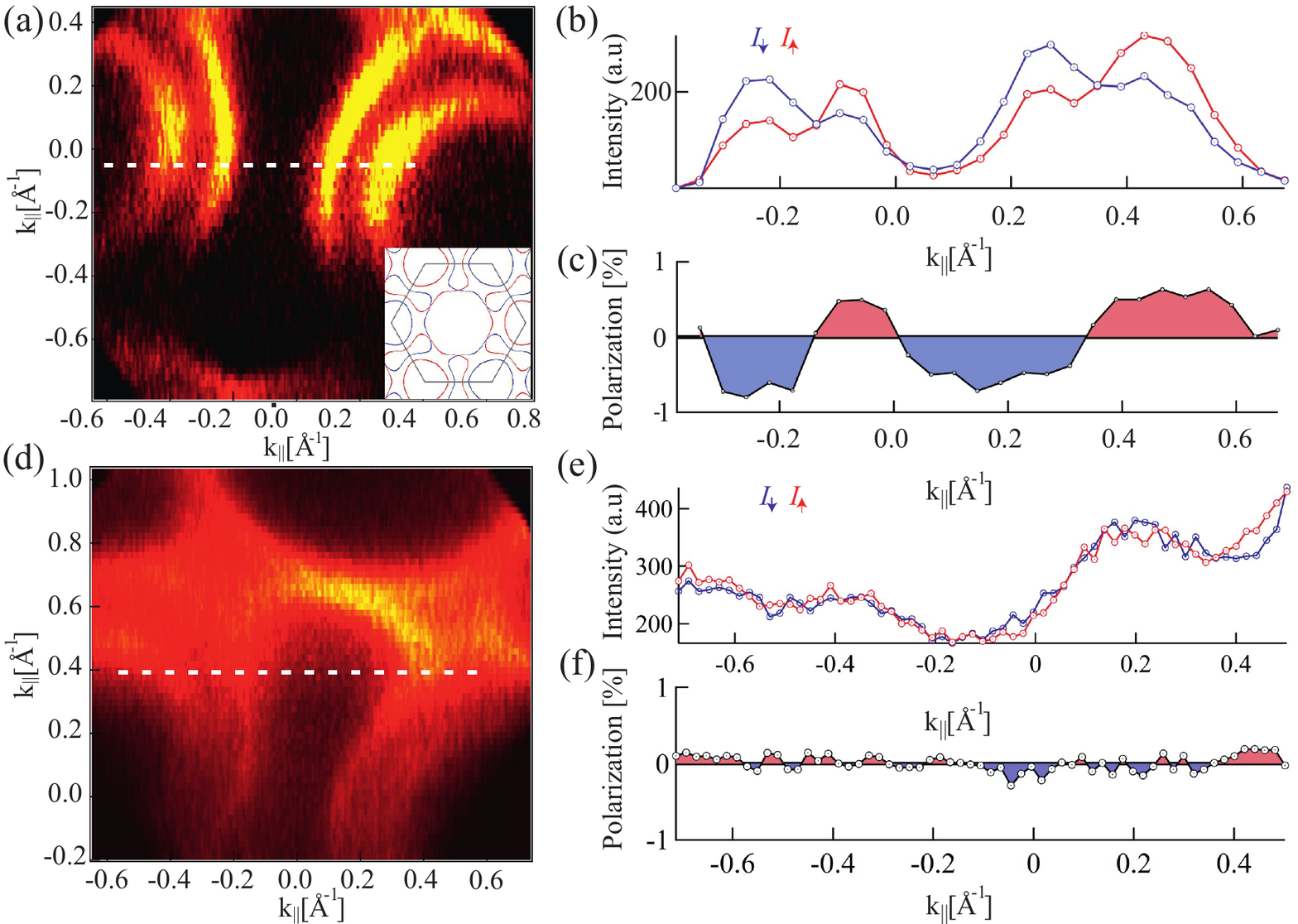}
	\caption{\textbf{Spin-valley locking.} \textbf{(a,d)}  A portion of the Fermi-surface of \TSB \textbf{(a)}\ and \TSH \textbf{(d)} measured using  $h\nu=$25~eV photons. The photon energy is optimized for maximum intensity of the polarized bands.  The white dashed lines represents the momentum cut for which the spin-resolved MDCs where measured. In both ARPES maps the (0,0) point was chosen to be the M point. \textbf{Inset}: DFT calculations of the spin polarized states derived from the 1H layer. $\textbf{(b)}$ Spin-polarized MDCs of \TSB measured at the Fermi-level  along the dashed line shown in \textbf{(a)}. The blue curve is the spin-down projection, and the red is spin up. Four peaks with different in testies are seen in each MDC, suggesting strong polarization. $\textbf{(c)}$ Polarization of the out-of-plane projection of the spin in \TSB, extracted from the MDCs in \textbf{(b)}.  $\textbf{(e, f)}$ Same as \textbf{(b)} and \textbf{(c)} for \TSH. No polarization is found in \TSH. }
	\label{Fig4}
\end{figure*}

Next, we estimate the effect of the 1T alternate stacking on the coupling between the 1H layers. A comparison between \TSH\ and \TSB\ reveals a completely different behaviors. In Fig. \ref{Fig2}(a,c), we present ARPES data measured along the K-M-K' line (k$_z$=0) for \TSH\ and \TSB, respectively. The black lines represents the DFT calculated dispersion. The gap between the inner and outer pockets at the M point in \TSH\ is a result of the inter-layer coupling between the 1H layers. In contrast, no gap is observed at the M-point in the \TSB\ spectra, similar to the mono-layer case \cite{Hoffman}.
The connection between the gap at the M-point and the inter-layer coupling was already pointed out \cite{xiao2012coupled,Yang2018enhanced}.

 The 1H monolayer exhibits an Ising spin texture due to the $M_z$ mirror symmetry. We use a simple effective model that includes two Ising layers to extract the inter-layer interaction  from the data:
\begin{equation}
H=\frac{\hbar^2}{2m^*}\sigma_0\tau_0k^2 + \alpha \sigma_z\tau_z k + \sigma_0 \tau_x t
\end{equation}
where $\sigma$ are the Pauli matrices for the spin and $\tau$ are the Pauli matrices representing the layer, $\alpha$ is the spin-orbit coupling (SOC) and t is the inter-layer coupling (ILC).
The best fit to the data (see supplementary Fig.~S4) yields an inter-layer coupling of  $65\pm5$~meV in \TSH\ and a vanishing inter-layer coupling in \TSB\, within our resolution. The Ising spin-orbit coupling  was set to be equal in both materials and is estimated to be $680\pm40$~meV\r{A}. Our results for \TSH\ agrees with previously measured values~\cite{de2018tuning}. 


Energy-distribution curves (EDCs) at different $k_z$ values were measured by varying the photon energy (see supplementary Fig.~S5). The EDCs along the M-L line for both polymorphs are shown in Fig. \ref{Fig2}(b,d). In \TSH (Fig. \ref{Fig2}(b)), we observe a significant dispersion with a bandwidth of about 150~meV. The band split, created by the inter-layer coupling, is maximal at the M-point and it vanishes at the L-point due to a glide-mirror symmetry that prevents the lift of the degeneracy~\cite{Bawden2016,dentelski2021robust}. 

In \TSB\ (Fig. \ref{Fig2}(d)), on the other hand, there is no dispersion at all, suggesting the inter layer coupling is negligible. We found no k$_z$ dependence in the width of the EDCs either, indicating that there is no measurable gap at any k$_Z$ point, as expected from a 2D system.

 The 2D character of \TSB\ can also explain the sharpness of the ARPES spectra at the Fermi-level compared to the data of \TSH. In a three-dimensional system, the finite k$_z$ resolution results in a broadening of the spectra.  


 At the Brillouin zone edge, it seems that the main role of the 1T layers is to reduce the inter-layer coupling. The effective mass of the valleys, estimated using our model, is $m^*=0.78\pm0.05~\mathrm{m_e}$ in both \TSB\ and \TSH. This suggests that the 1T layer mainly affects the band dispersion around the $\Gamma$ point. This is due to the in-plane nature of the orbitals forming the bands at the M and K points, which are $d_{xy}$ and $d_{x^2-y^2}$~\cite{gong2013magnetoelectric,Nayak2021evidence}. This is opposed to the orbitals at the $\Gamma$ point, $d_{z^2}$\cite{ritschel2015orbital,nayak2023first}, where hybridization between states of 1H and 1T layers takes place.

The negligible inter-layer coupling between the 1H layers in \TSB\ suggests a picture in which the electronic states are localized on individual 1H layers~\cite{xiao2012coupled,Zhang2014Spin,Riley2014,Bawden2016,Razzoli2017selective,xie2016manipulating,roch2019spin,huang2020hidden}. The combination of a large spin-orbit coupling with the non-centrosymmetric structure of the 1H layers is expected to result in a strong spin-valley locking in 1H-derived bands~\cite{Zhang2014Spin,Riley2014,Bawden2016,Razzoli2017selective} (see supplementary Fig.~S6). 

\begin{figure}[!ht]
	\centering
	\includegraphics[trim={0cm 5.5cm 0 0},clip, 
         width=0.8\linewidth]{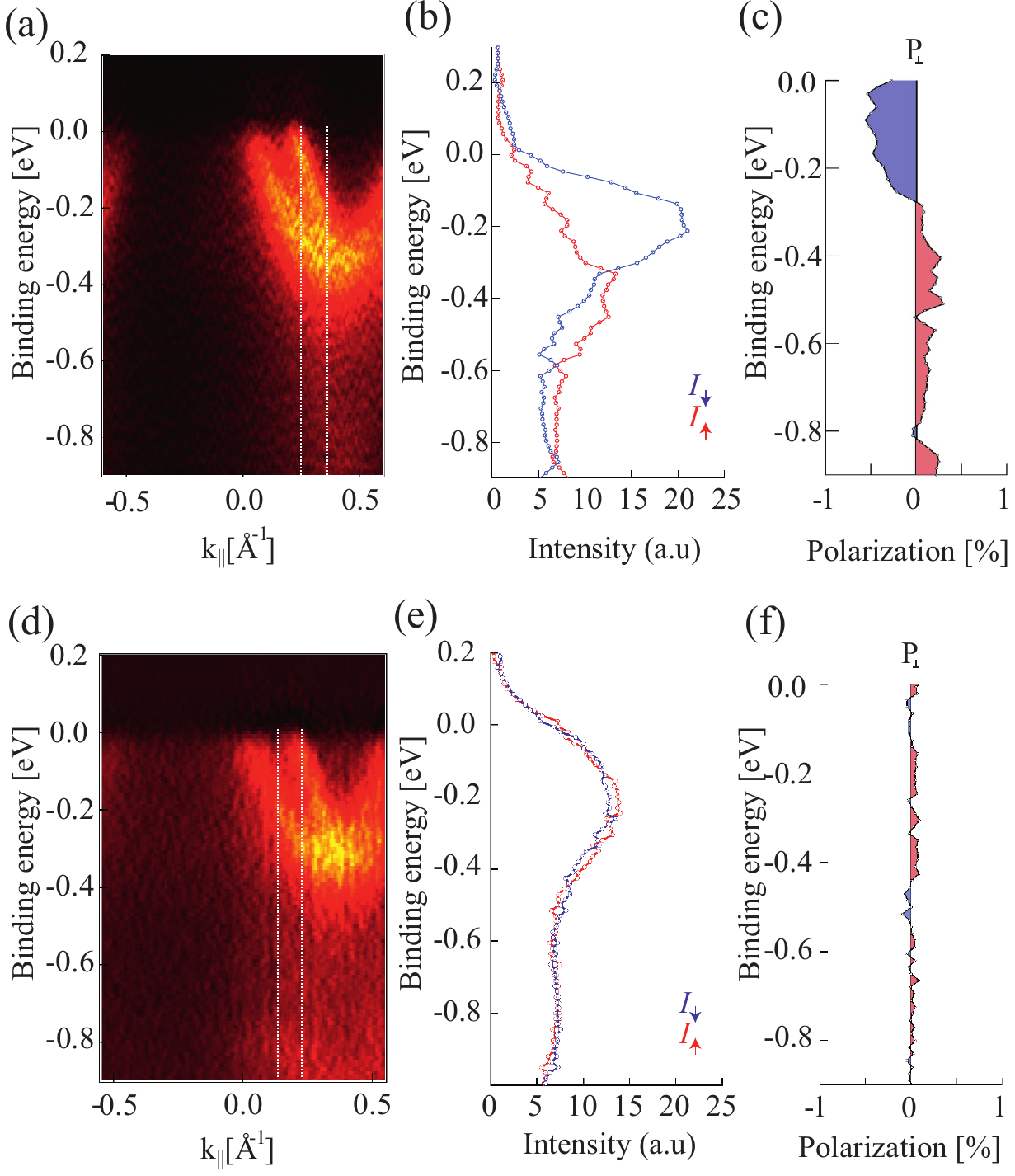}
     \caption{\textbf{ Valley polarization.} $\textbf{(a,d)}$ Band dispersion around the M-point for \TSB and \TSH, respectively, measured using 55eV photons. The dashed lines show the locations of the EDCs shown in \textbf{b} and \textbf{e} respectively. $\textbf{(b)}$ Spin-polarized EDCs cutting two bands in \TSB. The blue curve is the spin-down, and the red is spin-up, both normal to the surface. $\textbf{(c)}$ Spin polarization normal to plane extracted from the EDCs in \textbf{b}. A polarization of nearly 70\% is found for the band at low binding energy. For higher binding energies the polarization is weaker. $\textbf{(e, f)}$ Same as \textbf{b}  and \textbf{c} for \TSH. No polarization is found.}
	\label{Fig5}
\end{figure}

\subsection{Spin-Valley locking in 4Hb-TaS$_{2}$}

We use spin-resolved ARPES to directly measure the spin texture of the electronic bands and the Fermi-surface. 
First, we focus on the spin polarization at the Fermi-level. In Fig. \ref{Fig4}a, we present an ARPES intensity map of part of the Fermi-surface. One can observe the characteristic "dog-bone" shaped Fermi-pockets and the two electron pockets centered around K and K'. In the inset, we show the DFT calculated spin-resolved Fermi-surface. DFT finds perfect spin-valley locking, meaning that the electron pockets around K and K' have opposite polarization and the electron-spin polarization on the "dog-bone" changes on crossing the $\Gamma$-M line and remains the same when crossing the $\Gamma$-K line (see supplementary Fig.~S7). 
In Fig. \ref{Fig4}b, we show the spin-resolved MDCs measured along the dashed line in Fig. \ref{Fig4}a. The spin is measured normal to the plane, where the red and blue curves represent spin-up and spin-down, respectively.  The measured cut in momentum space crosses the Fermi-surface 4 times. From the MDCs it is clear that the polarization changes sign 4 times on each k$_F$ crossing. In Fig. \ref{Fig4}c, we show the spin polarization out of plane, defined as:  
   $ P = \frac{1}{S} \frac{I_{\uparrow}-I_{\downarrow}}{I_{\uparrow}+I_{\downarrow}}$ where $S$, is the Sherman function. 
We find a  polarization of up to 85\%, in good agreement with the DFT results.  
A measurement along a similar momentum cut in \TSH\ shows no polarization. The MDCs cross the same  4 bands (Fig.~\ref{Fig4}e), but the polarization shown in Fig. \ref{Fig4}f, is negligible.  

Next, we compare the polarization at different binding energies. ARPES spectra along the M-K line is shown in Figs.~\ref{Fig5}(a,d) for \TSB\ and \TSH, respectively. ARPES data were measured using 55~eV photons. In Fig. \ref{Fig5}(b,e), we present spin-resolved EDCs measured in the momentum range marked by white lines in Fig. \ref{Fig5}(a,d). The out-of-plane spin polarization is presented in Fig. \ref{Fig5}(c,f).
The same picture emerges; we find significant polarization in \TSB, while no polarization is found in \TSH. Additionally, we find that the polarization in \TSB\ is larger for low binding energies and decreases at high binding energies.

\section{Discussion}

We find that the presence of 1T layers between 2H layers has two main effects: 1) About 1 electron per star-of-David is transferred to the 1H layers, and 2) the 1T layers suppress the coupling between the 1H layers.

The charge transfer leads to an almost complete depletion of the 1T layers and it seems to rule out any Mott-like physics in \TSB, which is believed to be important in bulk \TST \cite{Lawson2014,law20171t}. Saying that, we note that we do find new small electron pockets around the $\Gamma$-point. Using STM, Kondo peaks were found in about 10\% of the Star-of-David sites, indicating that the charge from these sites was not transferred to the 1H layers \cite{nayak2023first}. It is possible that these sites are the origin of the small pockets we observe.  

Based on the ARPES data, it appears that there are almost no localized moments in the 1T layers, making it is difficult to understand how the remaining dilute charges in the 1T layers can account for the unconventional magnetism observed in the superconducting state of \TSB \cite{persky2022magnetic, liu2023magnetization}, nor for the breaking of time reversal symmetry \cite{Ribak2020}.


The diminishing inter-layer coupling has far reaching consequences. It suggests that \TSB\ can be thought of as a stack of 2D 1H superconducting layers separated by 1T layers. This can explain the increase in T$_c$. It is known that in \TSH\ T$_c$ increases as the samples become thinner, T$_c$ in the limit of a monolayer is similar to the T$_c$ of \TSB \cite{Yang2018enhanced}.

The absence of inter-layer coupling, combined with the strong spin-orbit coupling, naturally explains the Ising-superconductor behavior, as evident in the large anisotropy of  $\mathrm{H_{C2}}$ in \TSB \cite{Ribak2020}. The strong pinning of the spin out-of-plane results in an in-plane critical field  exceeding the Clogston-Chandrasekhar limit by a factor of 5.

Both \TSB\ and \TSH\ have an overall inversion symmetric-structure; nevertheless, we do observe strong valley-dependent polarization in \TSB. 
Each 1H layer is non-centrosymmetric and does allow for spin-valley locking. Overall the bands should be doubly degenerate and the spin-valley locking of each 1H layer should cancel with the contribution from the 180 deg rotated 1H' layer. However, since ARPES is a very surface-sensitive probe, it mainly probes the top-most 1H layer and the electrons from deeper layers will not cancel the net polarization. 

The level of polarization ARPES should detect depends mainly on how localized are the wave functions on the individual non-centrosymmetric 1H layers. A simple argument shows that the level of polarization of an individual layer is set by $P \propto \frac{1}{\sqrt{1+(t/\lambda_{soc})^2}}$ where $t$ and  $\lambda_{soc}$  are the inter-layer coupling and spin-orbit coupling parameters, respectively \cite{Zhang2014Spin}. Large inter-layer coupling leads to hybridization, and the individual layers are not spin polarized.  

The vanishing of the inter-layer coupling in \TSB\ naturally explains why we measure a large polarization in \TSB\ but it is not clear why we find no polarization in \TSH. Using the values for $t$ (65 $\pm$ 5~meV) and $\lambda_{soc}=\alpha \times k_F$ (130 $\pm$ 15~meV), we expect a polarization of about 90\% in \TSH.  
Moreover, we don't find polarization even at the L-point (data taken with 55eV photons), where inter-layer coupling is canceled due to a glide-mirror symmetry\cite{Bawden2016}.

Spin-valley locking was measured using ARPES in other TMDs  \cite{Riley2014,Bawden2016,Razzoli2017selective}.   Transport measurements in a few-layers of \TSH\ and 2H-NbSe$_2$, two very similar superconducting TMDs, found a smaller value of $t / \lambda_{soc}$ in \TSH \cite{de2018tuning}. Nevertheless, clear spin-valley locking was measured in 2H-NbSe$_2$ \cite{Bawden2016}, but not in \TSH.

A notable difference between the two materials is the absolute value of the inter-layer coupling, which is $\sim32$~meV for 2H-NbSe$_2$ and $\sim65$~meV for \TSH. This suggests that the polarization may not be exclusively determined by $t/ \lambda_{soc}$. 

There are experimental details related to the ARPES experiment that can reduce the measured polarization in \TSH\ compared to \TSB. First, the separation between the 1H layers in \TSB\ is double than in \TSH. For a given mean-free path of the photo-electrons, $L$, the average polarization of the two top-most layers is given by $\frac{1}{1+\exp{-\Delta L/L}}$, where $\Delta L $ is the inter-layer distance. For a mean-free path of 8 \AA, we get an average polarization of $\sim$ 0.8 for \TSB\ and $\sim$ 0.6 for \TSH. 

Second, larger bandwidth normal to the planes in \TSH, combined with the finite resolution in k$_z$, can lead to some broadening of the ARPES spectra that could mix the two spin-polarized bands. Based on the DFT calculation made for a 10-layers slab (see Supplementary material), we don't expect a large broadening as we move from the M-point. In addition, the spin-resolved measurement improves effectively the ability to resolved two overlapping bands. 

While, it seems unlikely that experimental details can mask completely a polarization of $\sim$ 60\%, it certainly could make it more difficult to observe the polarization in \TSH\ compared to \TSB.  

To summarize, we used ARPES to map the band structure of \TSB. Comparing the spectra of \TSB\ and \TSH\ we find two major differences. First, we observed a significant charge transfer from the 1T layers to the 1H layers. This charge transfer depletes almost completely the lower Hubbard band of the 1T layers, suggesting that the localized charges on the star-of-David cells that govern the physics of bulk \TST, play a much smaller role in the physics of \TSB. However, we do find new small shallow pockets around the $\Gamma$-point.  DFT shows that these new states reside mainly on the 1T-layers. The connection of these states with the unusual properties of \TSB\ is still not clear.

In addition, we find that the inter-layer coupling between 1H-layers in \TSB\ is effectively zero. This leads to strong spin-valley locking that is completely missing in bulk \TSH. The measured spin-texture can explain the Ising-superconductivity found in \TSB\ and may play a significant role in determining the superconducting order-parameter. 

We provide a full detailed description of the electronic structure of \TSB, it remains to understand how the unusual superconducting state emerges from this electronic structure.

 \section{Methods}
\subsection{Samples growth}
High quality single crystals of \TSH\ and \TSB\ were grown using the chemical vapor transport method, as described in \cite{DiSalvo1973preparation, Ribak2020}. Single crystals were grown using iodine as a transport agent in a sealed quartz, using a 3-zone furnace. The crystals have typical lateral size of a few mm and a thickness of about 0.5 mm. The samples are characterized by XRD, TEM, EDS and magnetization measurements (see supplementary Fig.~S8 and Fig.~S9). Se in 4Hb crystals is used to stabilize the 4Hb phase and improve the properties of the superconducting phase.  EDS measurements show that the actual amount of Se in the crystals is on the order of 0.2\% (See supplementary material). The composition of the sample used in this study is 4Hb-TaS$_{2-x}$Se$_{x}$, x$\sim$0.004.

\subsection{Angle and Spin resolved PES}
High resolution ARPES was done at the Ultra end-station at the SIS beam-line at PSI, Switzerland.  
Angle- and spin- resolved PES were performed at the RGBL-II beamline in BESSYII Berlin, Germany and the APE-LE beamline in Elettra, Italy. 
The light used at BESSY-II beamline is p polarized, while the light used at APE-LE is s polarized. Samples were cleaved in-situ and measured at temperatures ranging from 10 to 30 K. For the spin-resolved measurements, a VLEED detector was used in APE-LE, and Mott detectors at BESSYII. A Sherman function of S=0.33 was used to generate the measured spin polarization for the data from APE-LE, and S=0.16 for the data from BESSYII.

The bands polarization is given by~\cite{Bawden2016,Riley2014,Razzoli2017selective}:
\begin{equation}
    P = \frac{1}{S} \frac{I_{\uparrow}-I_{\downarrow}}{I_{\uparrow}+I_{\downarrow}}
\end{equation}

where P is the polarization measured and $\mathrm{I_{\uparrow}/I_{\downarrow}}$ is the measured intensity for each projection corrected by a relative detector efficiency calibration. 

To determine the k$_z$ dispersion from photon-energy-dependent ARPES, we use the free electron final state approximation for normal emission~\cite{Damascelli2003,Lahoud2013,Riley2014,almoalem2021link}:
\begin{equation}
    k_z = \sqrt{2m/\hbar^2 (E_{kin}+V_0)}
\end{equation}
where $\mathrm{V_0}$ is the inner potential and $\mathrm{E_{kin}}$ is the kinetic energy of the photo-emitted electron. Our photon energy range covers two complete Brillouin zones along k$_z$. Fitting the photon energies of the Brillouin zones center ($\Gamma$) and edge (Z) gives best agreement taking an inner potential of 9 eV. We use the same inner potential for both \TSH\ and \TSB.

\subsection{Density-functional theory}

We performed density-functional theory (DFT) calculations using the generalized gradient approximation (GGA) in the Perdew-Burke-Ernzerhof (PBE) form with the Vienna \textit{ab initio} package\cite{kresse1999ultrasoft}. 

The DFT-D2 method was used to account for van der Waals (vdW) interactions\cite{grimme2006semiempirical}. Spin-orbit coupling (SOC) was included in all calculations.
We employ a $\sqrt{13}\times\sqrt{13}\times1$ supercell as an initial approach to describe the CDW phase of 4Hb-TaS$_2$. 

To simulate the properties of 4Hb-TaS$_2$ in a computationally efficient manner, we simplify the system by considering a bilayer configuration of 1T-TaS$_2$ and 1H-TaS$_2$. These polytypes have been chosen to maintain the essential characteristics of the 4Hb phase. 

The Effective Band Structures (EBS) method in VASPKIT\cite{wang2021vaspkit} was used to directly compare with experimental measurements by unfolding the band structures of supercells into the Brillouin zone of primitive unit cells. 

To explore the surface states of 2H and 4Hb surfaces, we used 10-unitcell thick slab models and projected bands to the to H layer as the surface bands. Because we were interested in the H layer bands, we ignored the CDW distortion in the 1T layer in the 4Hb slab model use a 1$\times$1 unitcell in the basal plane. 

\section{Data availability}
The data that support the findings of this study are provided in the main text and the Supplementary Information. The original data are available from the corresponding author upon request.

\section{Acknowledgments}
 We acknowledge the Paul Scherrer Institute, Villigen, Switzerland for provision of synchrotron radiation beamtime at beamline SIS of the SLS.
 We acknowledge Elettra Sincrotrone Trieste for providing access to its synchrotron radiation facilities. We thank the Helmholtz-Zentrum Berlin for the allocation of synchrotron radiation beamtime.
 The work at the Technion was supported by Israeli Science Foundation grant number ISF-1263/21.
 This work has been partly performed in the framework of the nanoscience foundry and fine analysis (NFFA-MUR Italy Progetti Internazionali) facility.

\section{Author Contributions}
A.K. conceived the project. A.A. performed the ARPES measurments and analysed the data. R.G, Y.N., and I.M. helped with ARPES measurements. I.F. synthesized the single-crystals. J.K. performed DFT calculations. F.M., I.V., J.F., J.S.B, O.J.C., N.C.P. and M.S. provided support at the synchrotron beam-lines. B.Y. supervised the DFT calculations and performed theoretical modeling. A.A. and A.K. wrote the manuscript with contributions from all authors.

\section{Competing interests}
The Authors declare no Competing Financial or Non-Financial Interests.

\bibliographystyle{apsrev4-1}
\bibliography{SpinPolTaS2}

\end{document}